\documentclass[aps,showpacs,tightenlines,twocolumn,nofootinbib,nobibnotes,superscriptaddress]{revtex4-1}
\usepackage{amsmath,amssymb,amsfonts,bm}
\usepackage{graphicx}
\usepackage{epstopdf}
\usepackage{dcolumn}
\usepackage{mathrsfs}
\usepackage{tgtermes}
\usepackage[colorlinks=true,linkcolor=red,citecolor=blue, urlcolor=blue,bookmarks=false]{hyperref}
\bibpunct{[}{]}{,}{n}{}{}
\usepackage{verbatim}
\usepackage{amsthm,amsmath,amssymb}
\usepackage{mathrsfs}
\usepackage{booktabs}
\usepackage{multirow}

\begin{document}
\title{Four-dimensional Floquet topological insulator with an emergent second Chern number}
\date{\today}
\author{Zheng-Rong Liu}
\affiliation{Department of Physics, Hubei University, Wuhan 430062, China}
\author{Rui Chen}\email{chenr@hubu.edu.cn}
\affiliation{Department of Physics, Hubei University, Wuhan 430062, China}
\author{Bin Zhou}\email{binzhou@hubu.edu.cn}
\affiliation{Department of Physics, Hubei University, Wuhan 430062, China}
\affiliation{Key Laboratory of Intelligent Sensing System and Security of Ministry of Education, Hubei University, Wuhan 430062, China}

\begin{abstract}
Floquet topological insulators have been widely investigated in lower-dimensional systems. However, Floquet topological insulators induced by time-periodic driving in higher-dimensional systems remain unexplored. In this work, we study the effects of time-periodic driving in a four-dimensional (4D) normal insulator, focusing on topological phase transitions at the resonant quasienergy gap. We consider two types of time-periodic driving, including a time-periodic onsite potential and a time-periodic vector potential. We reveal that both types of time-periodic driving can transform the 4D normal insulator into a 4D Floquet topological insulator characterized by an emergent second Chern number.
Moreover, it is found that the topological phase of the 4D system can be modulated by tuning the strength and frequency of the time-periodic driving.
Our work will be helpful for the future investigation of Floquet topological insulators in higher dimensions.
\end{abstract}

\maketitle

\section{Introduction}
In the past decades, topological matter has become an important topic in condensed matter physics~\cite{RevModPhys.82.3045, RevModPhys.83.1057, RevModPhys.88.021004, RevModPhys.88.035005, RevModPhys.89.040502, RevModPhys.89.041004, W_lfle_2018, Shen_2017, Bernevig_2013}. Topological insulators (TIs) in $d$-dimensional space possess a $d$-dimensional gapped bulk and ($d-1$)-dimensional gapless boundary states~\cite{PhysRevLett.95.146802, PhysRevLett.95.226801, 10.1126/science.1133734, 10.1126/science.1148047, 10.1088/1367-2630/ac40cb}. Recently, four-dimensional (4D) TIs have attracted considerable attention due to recent advances in artificial metamaterials~\cite{Zhang_2001, PhysRevB.78.195424, Mochol_Grzelak_2018, doi:10.1126/science.aam9031, PhysRevLett.111.186803, PhysRevLett.129.196602, PhysRevResearch.2.023364, PhysRevB.108.085306}. The 4D TI hosts three-dimensional (3D) gapless boundary states, characterized by a topological invariant---the second Chern number~\cite{PhysRevB.78.195424, Mochol_Grzelak_2018, doi:10.1126/science.aam9031, PRXQuantum.2.010310, PhysRevB.108.085114, bouhon2023second}. The 4D TI cannot naturally arise in condensed matter systems due to the limited dimensionality. In artificial metamaterials, synthetic dimensions~\cite{PhysRevLett.108.133001, PhysRevLett.115.195303, PhysRevA.87.013814, PhysRevA.93.043827, PhysRevX.11.011016, 10.1093/nsr/nwac289, 10.1038/s41467-020-15940-3, 10.1093/nsr/nwaa065, PhysRevX.13.011003} and mapping 4D models onto lower-dimensional systems~\cite{10.1038/nature25000, 10.1038/nature25011, PhysRevLett.111.226401, PhysRevLett.109.135701, PhysRevB.98.094434, PhysRevB.98.125431, 10.1038/s41467-023-36359-6, 10.1038/s41467-023-36767-8} are promising schemes for implementing 4D TIs. Experimentally, the flexibility of atomic and photonic systems has inspired proposals to realize 4D topological physics~\cite{PhysRevLett.108.133001, PhysRevLett.115.195303, PhysRevA.87.013814, PhysRevA.93.043827, 10.1093/nsr/nwac289, 10.1515/nanoph-2022-0778}. Furthermore, since electric circuits are defined in terms of electronic elements and their interconnections, lattices with genuine 4D structures can be constructed by applying appropriate capacitive and inductive connections~\cite{10.1038/s41467-020-15940-3, 10.1093/nsr/nwaa065, 10.1038/s41467-023-36359-6, 10.1038/s41467-023-36767-8}.

Floquet engineering is a controlled protocol to induce or manipulate exotic topological properties by time-periodic driving~\cite{PhysRevLett.114.246802, PhysRevB.96.020507, doi:10.1126/science.1239834, 10.1038/nmat4156, 10.1038/nphys3609, 10.1038/s41567-019-0698-y, 10.1038/s41467-019-12231-4, 10.1038/s41586-023-05850-x, 10.1038/s41586-022-05610-3, PhysRevLett.110.016802, PhysRevB.98.235159, PhysRevB.99.075121, PhysRevB.101.075108, PhysRevB.103.L100301, PhysRevLett.121.196401, PhysRevLett.121.237401, PhysRevB.105.L161108, PhysRevB.106.235405, PhysRevB.108.L020303, PhysRevB.108.075435, PhysRevX.3.031005}. Time-periodic driving can induce topologically nontrivial matter in trivial static systems, and such topological matter is known as Floquet topological matter, include Floquet topological insulators (FTIs)~\cite{10.1038/nphys1926, PhysRevB.79.081406, PhysRevB.82.235114, PhysRevLett.107.216601, 10.1038/nature12066, 10.1002/pssr.201206451, 10.1038/s42254-020-0170-z}, Floquet topological semimetals~\cite{PhysRevLett.120.237403, 10.1209/0295-5075/105/17004, PhysRevE.93.022209, PhysRevB.93.144114, PhysRevB.94.075443, PhysRevB.97.155152, PhysRevB.99.115136, PhysRevResearch.2.033045, PhysRevB.107.L121407}, etc. In lower-dimensional ($d\leq3$) systems, considerable research on the FTI has been reported, such as the discovery of many intriguing topological phases that are absent in static systems~\cite{PhysRevLett.113.266801, PhysRevLett.121.036401, PhysRevLett.123.016806, PhysRevLett.124.057001, PhysRevLett.124.216601, PhysRevA.100.023622, PhysRevB.93.184306, PhysRevB.101.174314, PhysRevB.103.115308, PhysRevB.106.184106, PhysRevB.108.L241402, 10.1038/s42254-020-0170-z}. Up to now, FTIs have been realized by experiments for solid-state~\cite{doi:10.1126/science.1239834, 10.1038/s41567-019-0698-y, 10.1364/prj.404163}, photonic~\cite{10.1038/ncomms1872, 10.1038/nature12066, 10.1038/ncomms13756, 10.1038/ncomms13918, RevModPhys.91.015006, 10.1038/s41563-020-0641-8}, acoustic~\cite{10.1038/ncomms11744, 10.1038/ncomms13368}, electric circuits~\cite{dabiri2023electric}, and ultracold atom systems~\cite{10.1038/nature13915, 10.1038/s41567-019-0417-8, 10.1038/s41567-020-0949-y}. Then a question naturally arises whether the FTI phase can occur in 4D systems.

In this paper, we answer the above question and provide the scheme for inducing the 4D FTI via two types of time-periodic driving. In the 4D FTI, the 4D bulk is characterized by the second Chern number, accompanied by the 3D boundary states. First, we study the influence of the time-periodic onsite potential ${\boldsymbol{V}}(\tau)$ on the 4D system. When the frequency of the time-periodic onsite potential $\omega$ is smaller than the bandwidth of the static system $E_{W}$, the time-periodic onsite potential can induce a phase transition from the trivial static system to a 4D TI. This 4D TI possesses 3D gapless boundary states, characterized by a nonzero second Chern number $C_{2}=-3$, dubbed the 4D FTI. Notably, when the amplitude of the time-periodic onsite potential $V$ exceeds a critical value, we find that the time-periodic onsite potential can transform a topologically nontrivial phase with $C_{2}=-3$ to another topologically nontrivial phase with $C_{2}=3$. Additionally, our investigation reveals that the time-periodic onsite potential with different frequency can induce a topological phase transition in a 4D normal insulator, leading to the emergence of a 4D FTI with distinct Chern numbers, namely $C_{2}=-3$, $C_{2}=-1$, or $C_{2}=3$. Second, we demonstrate that the time-periodic vector potential ${\boldsymbol{A}}(\tau)$ can induce the emergence of the 4D FTI with $C_{2}=2$. However, when the amplitude $A$ exceeds a critical value, the time-periodic vector potential destroys the topological properties of the Floquet system, accompanied by the decay of second Chern number from $C_{2}=2$ to $C_{2}=0$.

The rest of the paper is organized as follows. In Sec.~\ref{SecII.A}, we introduce a time-periodic onsite potential in the 4D Dirac model and demonstrate the method for calculating the second Chern number. Then, we present a 4D FTI driven by the time-periodic onsite potential in Sec.~\ref{SecII.B}. In Sec.~\ref{SecIII}, we investigate the topological phase transition of the 4D FTI driven by the time-periodic vector potential. Finally, we summarize our conclusions in Sec.~\ref{Conclusion}.

\section{Time-periodic onsite potential}
\label{SecII}

\subsection{Model}
\label{SecII.A}
We first study the influence of the time-periodic onsite potential on the 4D system. The time-dependent 4D TI model is given by the following equation:
\begin{align}
H(\textbf{k},\tau)=H(\textbf{k})+{\boldsymbol{V}}(\tau).
\label{Eq1}
\end{align}
The first term is a static Hamiltonian describing the 4D TI~\cite{PhysRevB.78.195424},
\begin{align}
H(\textbf{k})=&\sin(k_{x})\Gamma_{2}+\sin(k_{y})\Gamma_{3}+\sin(k_{z})\Gamma_{4}+\sin(k_{w})\Gamma_{5}\nonumber\\
&+m(\textbf{k})\Gamma_{1},
\end{align}
where the Dirac matrices $\Gamma_{j}=(\sigma_{x}\otimes \sigma_{0}, \sigma_{y}\otimes \sigma_{0}, \sigma_{z}\otimes \sigma_{x}, \sigma_{z}\otimes \sigma_{y}, \sigma_{z}\otimes \sigma_{z}), j=1, 2, 3, 4, 5,$ satisfying the anticommutation relations $\{\Gamma_{i},\Gamma_{j}\}=2\delta_{ij}$. $m(\textbf{k})=m+c[\cos(k_{x})+\cos(k_{y})+\cos(k_{z})+\cos(k_{w})]$, $m$ is the Dirac mass, and $c$ denotes the nearest-neighbour hopping amplitude. In subsequent calculations, $c=1$. The second term represents the time-periodic onsite potential ${\boldsymbol{V}}(\tau)=V\cos(\omega\tau)\Gamma_{1}$, where $V$ is the amplitude of the time-periodic onsite potential, and $\omega$ is the frequency.

In four dimensions, the response of the current to an electric field $\textbf{E}$ and a magnetic field $\textbf{B}$ is related to two topological invariants~\cite{RevModPhys.83.1057, PhysRevB.78.195424, Mochol_Grzelak_2018}:
\begin{align}
j_{l}=\frac{e^{2}}{h}\sum_{\varepsilon_{\alpha}<\varepsilon_{F}}\frac{1}{(2\pi)^{4}}E_{m}\int \Omega_{lm}^{\alpha}d^{4}k+\frac{C_{2}(\varepsilon_{F})}{4{\pi^{2}}}\epsilon_{lmno}E_{m}B_{no}.
\end{align}
The quantized linear response is associated with the first Chern number, and the quantized nonlinear response of current to the electric field and the magnetic field is correlated with the second Chern number $C_{2}$. The second Chern number is given by the following formula~\cite{PhysRevB.78.195424, Mochol_Grzelak_2018, doi:10.1126/science.aam9031, PRXQuantum.2.010310, PhysRevB.108.085114}:
\begin{align}
C_{2}=\frac{1}{4\pi^{2}}\int d\textbf{k}\text{Tr}[\Omega_{xy}\Omega_{zw}+\Omega_{wx}\Omega_{zy}+\Omega_{zx}\Omega_{yw}],
\end{align}
with the non-Abelian Berry curvature
\begin{align}
\Omega_{mn}^{\alpha\beta}=\partial_{m}a_{n}^{\alpha\beta}-\partial_{n}a_{m}^{\alpha\beta}+i[a_{m},a_{n}]^{\alpha\beta},
\end{align}
where $m, n=x, y, z, w$, and the Berry connection of the occupied bands $a_{m}^{\alpha\beta}=-i\left\langle u^{\alpha}(\textbf{k})\right\vert\frac{\partial}{\partial k_{m}}\left\vert u^{\beta}(\textbf{k})\right\rangle$, $\left\vert u^{\alpha}(\textbf{k})\right\rangle$ denotes the occupied eigenstates below the Fermi energy $\varepsilon_{F}$ with $\alpha=1, \dots, N_{\rm{occ}}$. When the amplitude of the time-periodic onsite potential $V=0$, the static system exhibits different topological phases for different Dirac mass:
\begin{align}
\begin{split}
C_{2}(m)=
\begin{cases}
0, &m<-4\\
1, &-4<m<-2\\
-3, &-2<m<0\\
3, &0<m<2\\
-1, &2<m<4\\
0, &m>4.
\end{cases}
\end{split}
\end{align}
The values of the second Chern number $|C_{2}|$ imply the number of 3D gapless boundary states for the 4D system.

Now we investigate the effect of time-periodic onsite potential on the 4D system. Floquet theory is often used to study time-dependent systems~\cite{PhysRevB.96.054207}. Based on the Floquet theory, we can convert time-dependent Hamiltonian $H(\tau)$ into time-independent Floquet Hamiltonian $H_{F}$ by using the Fourier transformation. The Floquet Hamiltonian $H_{F}$ is an infinite Hamiltonian with the following form:
\begin{align}
H_{F}=\begin{pmatrix}
\ddots & \vdots & \vdots & \vdots & \ddots \\
\cdots & H_{0}-\omega & H_{+1} & H_{+2} & \cdots \\
\cdots & H_{-1} & H_{0} & H_{+1} & \cdots \\
\cdots & H_{-2} & H_{-1} & H_{0}+\omega  & \cdots \\
\ddots & \vdots & \vdots & \vdots & \ddots
\end{pmatrix},
\end{align}
where
\begin{align}
H_{l}&=\frac{1}{T}\int_{0}^{T}d\tau H(\tau)e^{i l\omega\tau},
\end{align}
$\omega$ and $T=\frac{2\pi}{\omega}$ represent the frequency and period of the time-periodic onsite potential, respectively. $l$ can be taken as $0$, $\pm1$, $\pm2$, $\cdots$. After applying the Fourier transformation to the time-dependent 4D TI model $H(\textbf{k},\tau)$ [Eq.~(\ref{Eq1})], the block matrices of the Floquet Hamiltonian $H_{F}$ are shown below:
\begin{align}
H_{0}=&\sin(k_{x})\Gamma_{2}+\sin(k_{y})\Gamma_{3}+\sin(k_{z})\Gamma_{4}+\sin(k_{w})\Gamma_{5}\nonumber\\
&+\left [ m+c\sum_{n=x,y,z,w}\cos(k_{n}) \right ] \Gamma_{1},\nonumber\\
H_{-1}=&H_{+1}^{\dagger}=\frac{V}{2}\Gamma_{1},\nonumber\\
H_{|l|\geq2}=&0.
\end{align}
In subsequent calculations, the infinite Floquet Hamiltonian is truncated when the results are convergent, and the Fermi energy $\varepsilon_{F}$ is set to $\omega/2$. In subsequent calculations, the second Chern number refers to $C_{2}(\varepsilon_{F}=\omega/2)$.

\begin{figure}[t]
	\includegraphics[width=8.5cm]{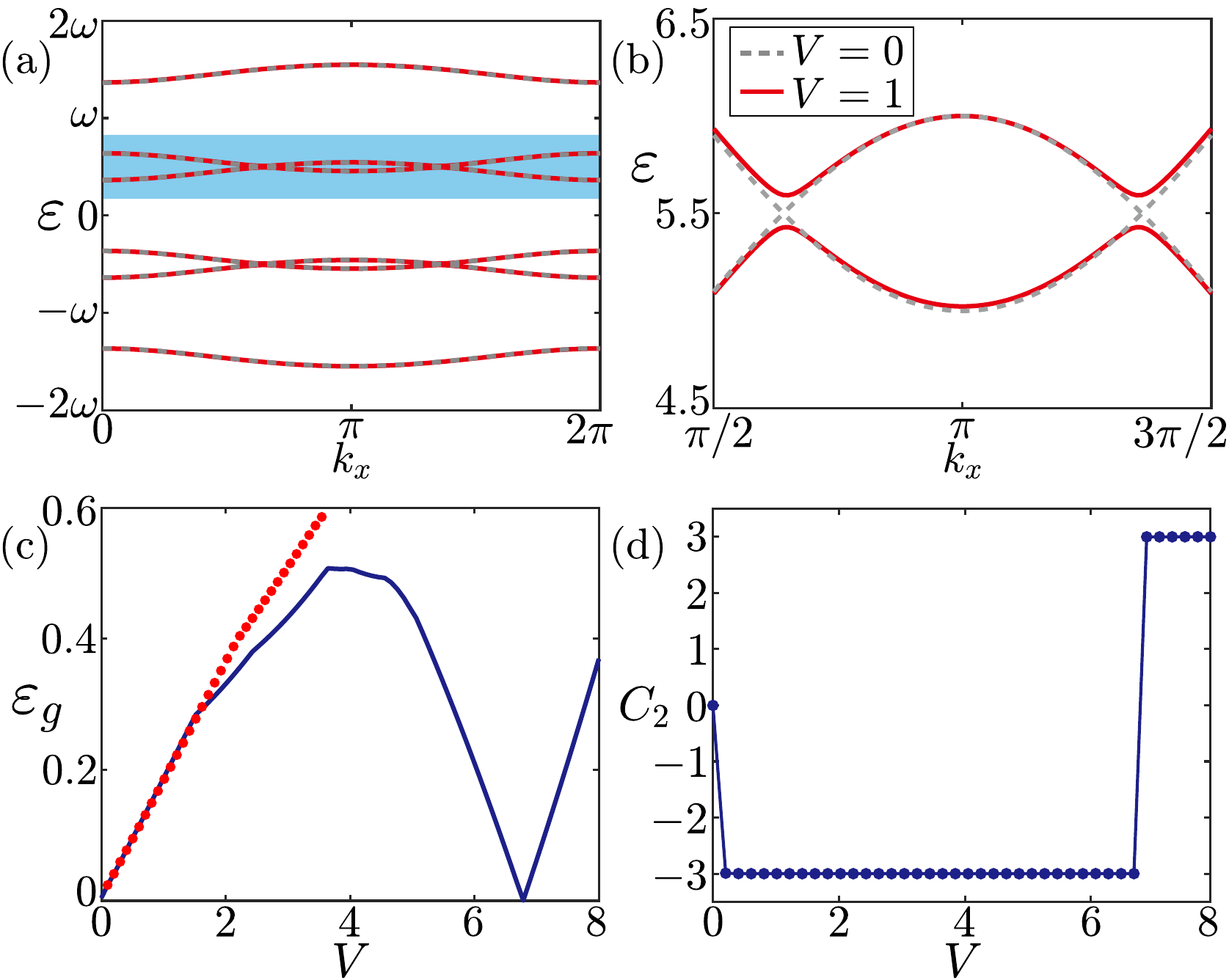} \caption{(a) The bulk quasienergy spectra for the Floquet Hamiltonian $H_{F}$ when $k_{y}=\pi, k_{z}=0, k_{w}=0$. (b) shows the quasienergy spectra of the cyan region in (a). The gray dashed line (red solid line) represents the amplitude of the time-periodic onsite potential $V=0$ ($V=1$). (c) The bulk gap $\varepsilon_{g}$ and (d) the second Chern number $C_{2}$ as a function of the amplitude of the time-periodic onsite potential $V$. Those red solid dots in (c) represent the bulk gap of the effective Hamiltonian $H_{\rm{RWA}}$ obtained by the rotating wave approximation. Here, we choose $\omega=11$ and $m=-6$.}%
	\label{fig1}
\end{figure}

\subsection{Periodic-driven 4D FTI}
\label{SecII.B}
In the trivial phase ($m=-6$), the static system hosts a trivial bulk gap and the bandwidth of the system is $E_{W}=2|m|+8=20$. We can study the effect of time-periodic onsite potential ${\boldsymbol{V}}(\tau)$ on the 4D trivial system by solving the Floquet Hamiltonian $H_{F}$, $H_{F}\left\vert\psi^{\alpha}\right\rangle=\varepsilon_{\alpha}\left\vert\psi^{\alpha}\right\rangle$. In the Floquet Hamiltonian $H_{F}$, the diagonal block $H_{0}\pm\omega$ is a copy of the original block $H_{0}$, shifting in energy by $\omega$. The band of the diagonal block $H_{0}$ ($H_{0}\pm\omega$) is referred to as the undriven (driven) band. As shown by the gray dashed line in Figs.~\ref{fig1}(a) and \ref{fig1}(b), the driven and undriven bands of the Floquet Hamiltonian $H_{F}$ cross each other when $\omega<E_{W}$. After turning on the amplitude of the time-periodic onsite potential $V$, the off-diagonal blocks $H_{l}(l\neq0)$ of the Floquet Hamiltonian hybridize the resonant quasienergies and gap them out, as shown by the red solid line in Fig.~\ref{fig1}(b).

\begin{figure}[t]
	\includegraphics[width=7.5cm]{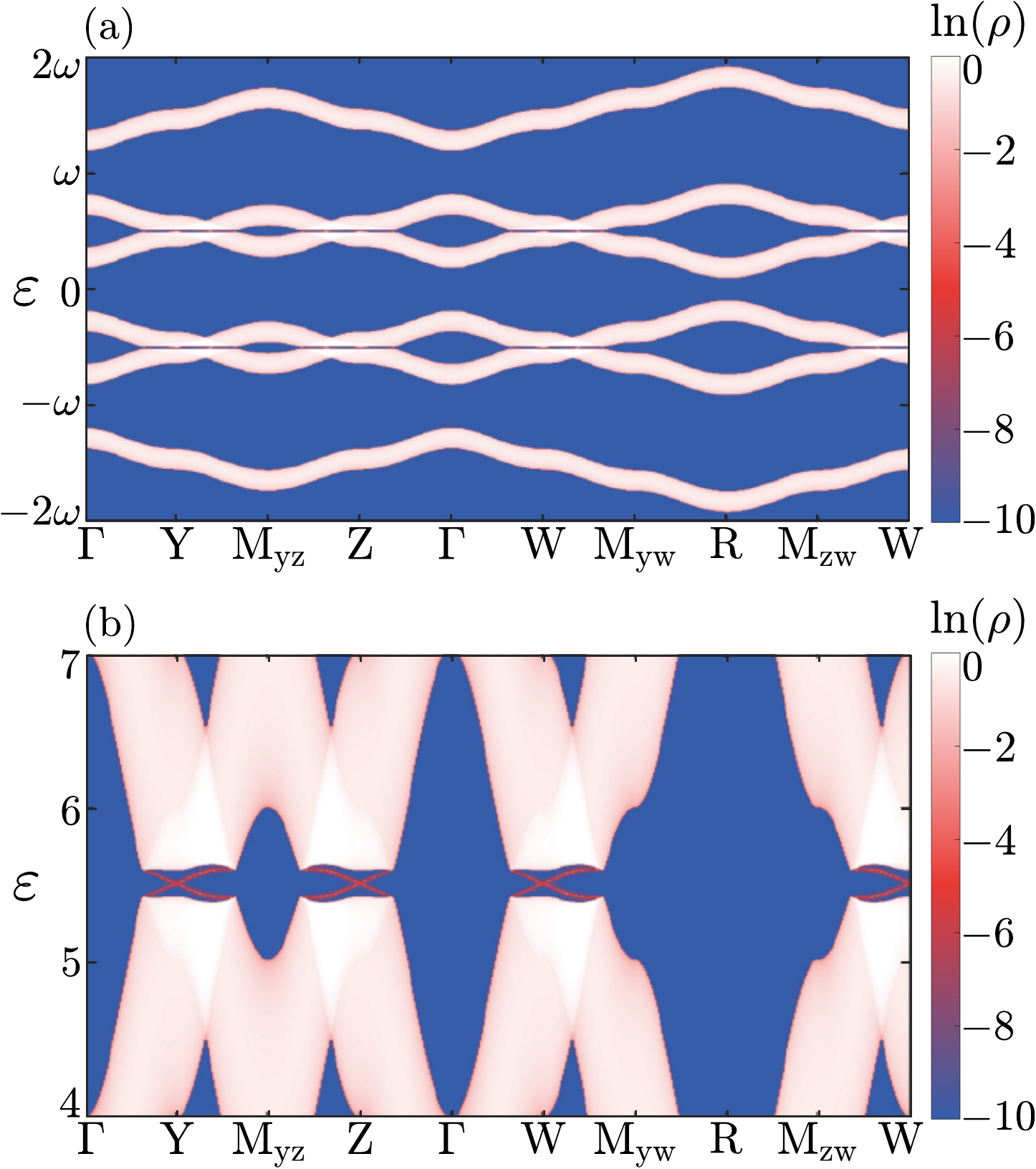} \caption{(a) Band structure in the plane $(\varepsilon, \textbf{k})$ when the open boundary condition is along the $x$ direction. The indicators of the horizontal axis are the high symmetry $\textbf{k}$ points in the first Brillouin zone of the quasi-3D system, $\Gamma=(k_{y}=0, k_{z}=0, k_{w}=0)$, ${\rm{Y}}=(\pi, 0, 0)$, ${\rm{Z}}=(0, \pi, 0)$, ${\rm{W}}=(0, 0, \pi)$, ${\rm{M_{yz}}}=(\pi, \pi, 0)$, ${\rm{M_{yw}}}=(\pi, 0, \pi)$, ${\rm{M_{zw}}}=(0, \pi, \pi)$, ${\rm{R}}=(\pi, \pi, \pi)$. (b) Band structure in the quasienergy interval $\varepsilon\in(4, 7)$ in which there are gapless boundary states. Here, we choose $\omega=11$, $V=1$, and $m=-6$.}%
	\label{fig2}
\end{figure}

In order to explore the phase transition of the 4D Floquet system, we calculate the bulk gap in the resonant quasienergy region as a function of the amplitude $V$ as shown in Fig.~\ref{fig1}(c). When $V=0$, the undriven conduction bands and the driven valence bands mix with each other in the resonant quasienergy region; thus, the bulk gap of the system is zero. This bulk gap can be opened by the time-periodic onsite potential with $V>0$. In Fig.~\ref{fig1}(d), we show the second Chern number as a function of $V$, and one can find that the bulk gap opened by the time-periodic onsite potential in the resonant quasienergy region is topologically nontrivial. In the interval $V\in(0, V_{c}\approx6.785)$, the second Chern number of the system maintains a quantized plateau $C_{2}=-3$. When $V=V_{c}$, the bulk gap closes at points $K_{1}=(k_{x}=\pi,k_{y}=\pi,k_{z}=0,k_{w}=0)$, $K_{2}=(\pi,0,\pi,0)$, $K_{3}=(\pi,0,0,\pi)$, $K_{4}=(0,\pi,\pi,0)$, $K_{5}=(0,\pi,0,\pi)$, and $K_{6}=(0,0,\pi,\pi)$. When $V>V_{c}$, the bulk gap reopens and the system transitions from a topologically nontrivial phase with $C_{2}=-3$ to another topologically nontrivial phase with $C_{2}=3$.

In Figs.~\ref{fig2}(a) and \ref{fig2}(b), we show the quasienergy spectra of the system when the open boundary condition is along the $x$ direction. The color bar represents the natural logarithm of the density of states ${\rm{ln}}(\rho)$, where $\rho$ is normalized to one. It can be found that there are three gapless Dirac points in the resonant quasienergy gap and no Dirac points in the gap near $\varepsilon=0$. These Dirac points are distributed at the high symmetry points ${\rm{Y}}=(k_{y}=\pi, k_{z}=0, k_{w}=0)$, ${\rm{Z}}=(0, \pi, 0)$, and ${\rm{W}}=(0, 0, \pi)$. The system with three gapless Dirac points is dubbed 4D FTI, characterized by a nonzero second Chern number $C_{2}=-3$. Moreover, in another topologically nontrivial phase with $C_{2}=3$, there are three gapless Dirac points distributed at ${\rm{M_{yz}}}=(k_{y}=\pi, k_{z}=\pi, k_{w}=0)$, ${\rm{M_{yw}}}=(\pi, 0, \pi)$, and ${\rm{M_{zw}}}=(0, \pi, \pi)$ when the open boundary condition is along the $x$ direction.

\begin{figure*}[t]
	\includegraphics[width=14cm]{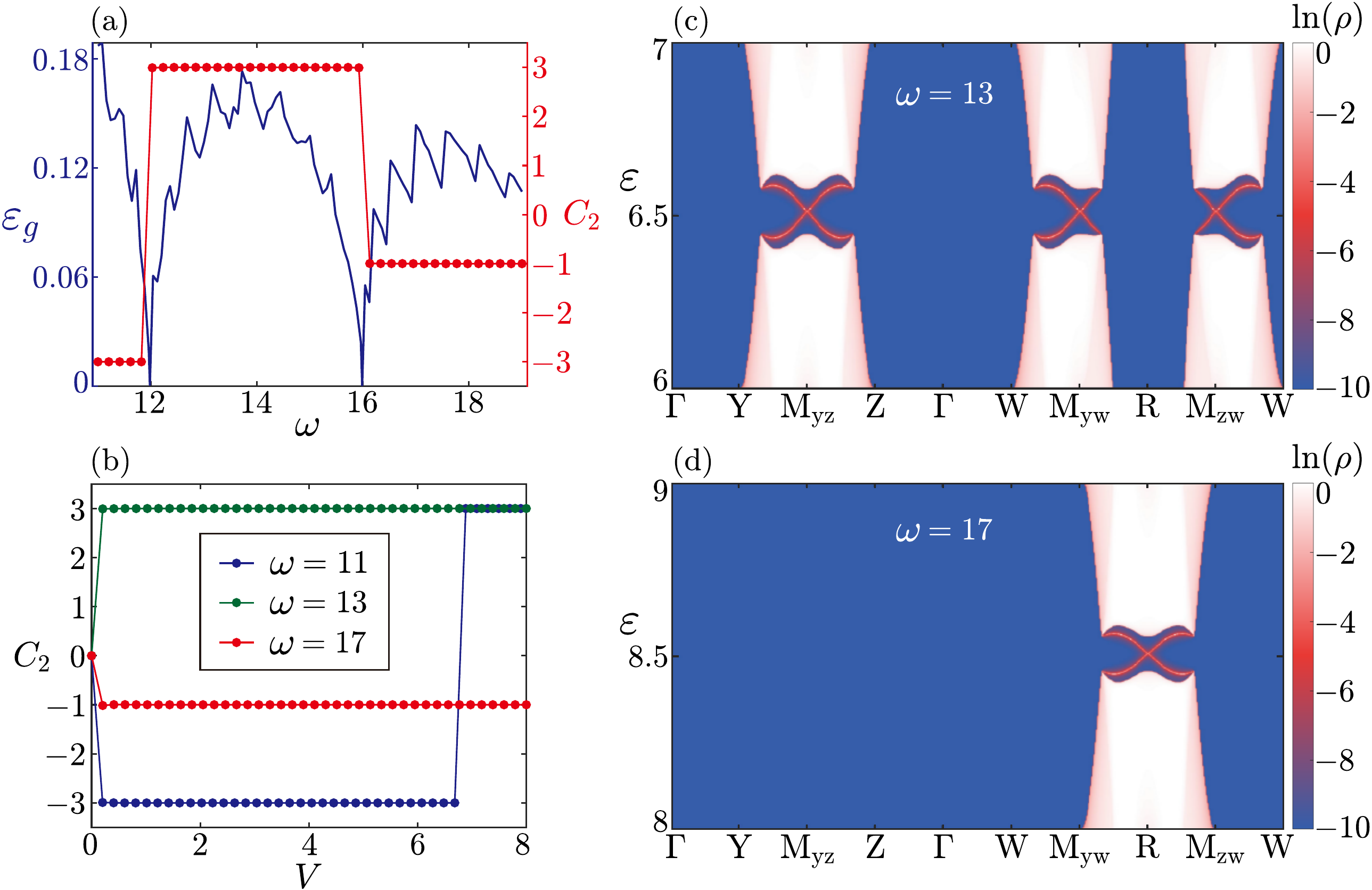} \caption{(a) The bulk gap $\varepsilon_{g}$ and the second Chern number $C_{2}$ as a function of the frequency $\omega$ when $V=1$. (b) The second Chern number $C_{2}$ as a function of $V$.  The blue, green, and red dotted lines correspond to $\omega=11$, $\omega=13$, and $\omega=17$, respectively. Band structure of the resonant quasienergy region in the plane $(\varepsilon, \textbf{k})$ under the open boundary condition along the $x$ direction for (c) $\omega=13$ and (d) $\omega=17$. The indicators of the horizontal axis are the high symmetry $\textbf{k}$ points in the first Brillouin zone of the quasi-3D system, $\Gamma=(k_{y}=0, k_{z}=0, k_{w}=0)$, ${\rm{Y}}=(\pi, 0, 0)$, ${\rm{Z}}=(0, \pi, 0)$, ${\rm{W}}=(0, 0, \pi)$, ${\rm{M_{yz}}}=(\pi, \pi, 0)$, ${\rm{M_{yw}}}=(\pi, 0, \pi)$, ${\rm{M_{zw}}}=(0, \pi, \pi)$, ${\rm{R}}=(\pi, \pi, \pi)$. In (c) and (d), $V=1$. And in all plots, we choose $m=-6$.}%
	\label{fig3}
\end{figure*}

Furthermore, we show the bulk gap $\varepsilon_{g}$ and the second Chern number $C_{2}$ as a function of the frequency $\omega$ when $V=1$ in Fig.~\ref{fig3}(a). The solid blue line and the red dotted line represent the bulk gap and the second Chern number, respectively. It can be found that the topological phase of the 4D FTI can be transformed by tuning the frequency of the time-periodic onsite potential. When $\omega\approx11.98$, the bulk gap of the 4D system closes, accompanied by a topological phase transition from $C_{2}=-3$ to $C_{2}=3$. And when $\omega\approx15.985$, the system transitions from a topological phase with $C_{2}=3$ to another topological phase with $C_{2}=-1$. Simultaneously, Figure~\ref{fig3}(b) illustrates the evolution of the second Chern number $C_{2}$ with the amplitude $V$ under the influence of the time-periodic driving with different frequencies in the 4D system. It is evident that the time-periodic driving with different frequencies can transform a 4D normal insulator into a 4D Floquet topological insulator with distinct second Chern numbers. In Figs.~\ref{fig3}(c) and \ref{fig3}(d), we present the band structures under the open boundary condition along the $x$ direction for $\omega=13$ and $\omega=17$, respectively. When $\omega=13$, the topologically nontrivial system possesses three gapless Dirac points distributed at ${\rm{M_{yz}}}=(k_{y}=\pi, k_{z}=\pi, k_{w}=0)$, ${\rm{M_{yw}}}=(\pi, 0, \pi)$, and ${\rm{M_{zw}}}=(0, \pi, \pi)$, characterized by the second Chern number $C_{2}=3$. And when $\omega=17$, the topologically nontrivial system possesses a single gapless Dirac point distributed at ${\rm{R}}=(\pi, \pi, \pi)$, characterized by the second Chern number $C_{2}=-1$.

In the limit where the frequency $\omega$ is (nearly) resonant with the level spacing of the Hamiltonian, we can apply the rotating wave approximation (RWA) to analyze the time-dependent system~\cite{10.1038/nphys1926, PhysRevB.96.054207, ZEUCH2020168327}. For the time-dependent Hamiltonian $H(\textbf{k},\tau)=H(\textbf{k})+V\cos(\omega\tau)\Gamma_{1}$, we can convert it to the rotating frame by using the unitary transformation,
\begin{align}
H_{\rm{rot}}=U^{\dagger}\left [ H(\textbf{k},\tau)-i\frac{\partial}{\partial\tau} \right ] U,
\end{align}
where $U=P_{+}+P_{-}e^{i\omega\tau}$, and $P_{\pm}$ are projectors on the unoccupied and occupied bands of $H(\textbf{k})$. In the rotating-frame Hamiltonian $H_{\rm{rot}}$, we can neglect the fast-oscillating terms as they quickly average to zero~\cite{10.1038/nphys1926, PhysRevB.96.054207}. Hence, we find a time-independent Hamiltonian $H_{\rm{RWA}}$ as follows:
\begin{align}
H_{\rm{RWA}}=H(\textbf{k})+\omega P_{-}P_{-}+\frac{V}{2}\left ( P_{+}\Gamma_{1}P_{-}+P_{-}\Gamma_{1}P_{+} \right ).
\end{align}
In Fig.~\ref{fig1}(c), we show the bulk gap of the Hamiltonian $H_{\rm{RWA}}$ as a function of $V$, labeled by red solid dots. When the amplitude of the time-periodic onsite potential $V$ is small, the results obtained by the rotating wave approximation coincide with those obtained by solving the Floquet Hamiltonian $H_{F}$.

\section{Time-periodic vector potential}
\label{SecIII}
\begin{figure}[t]
	\includegraphics[width=8.5cm]{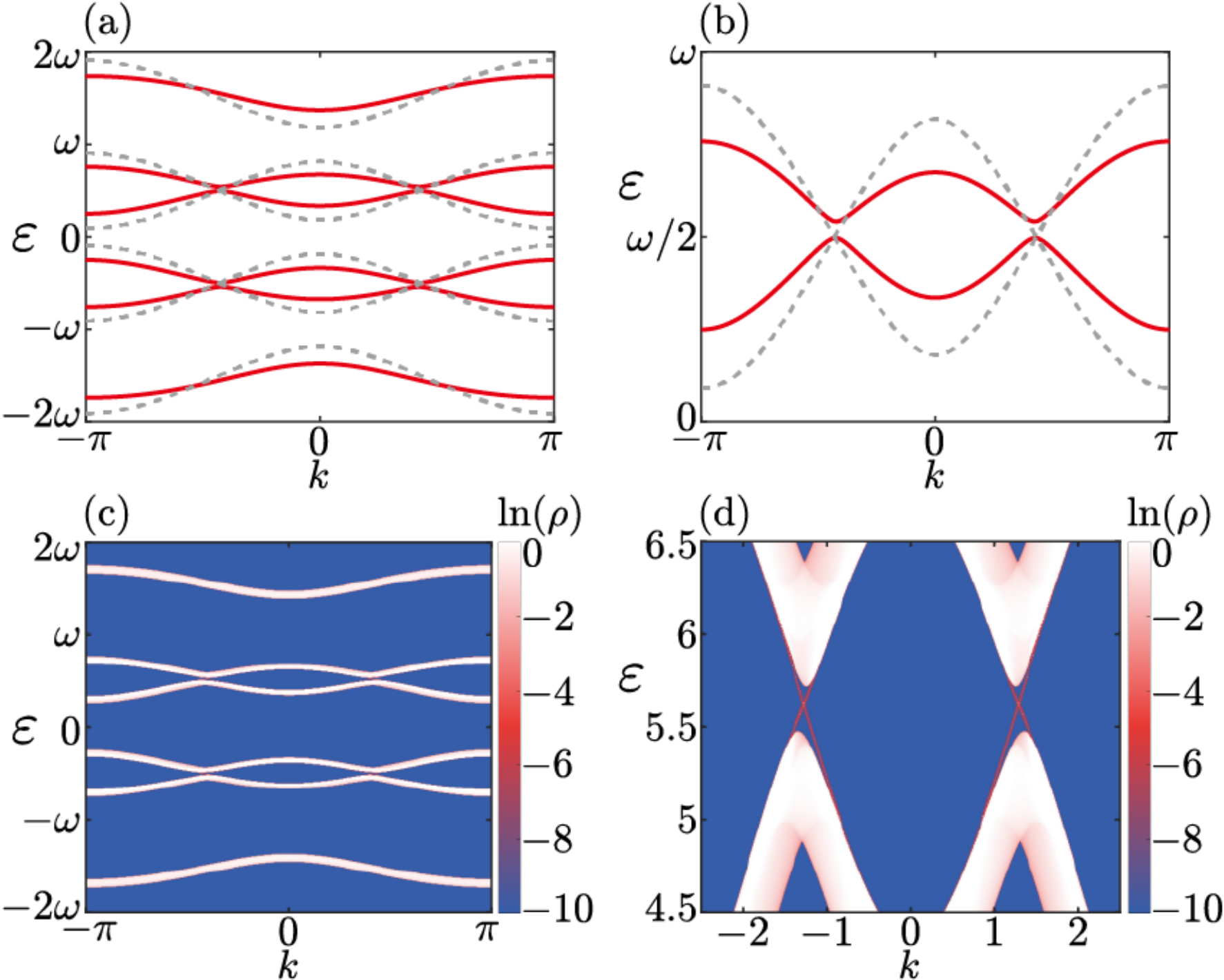} \caption{(a) The bulk quasienergy spectra for the Floquet Hamiltonian $\mathcal{H}_{F}$ when $k=k_{n}$ ($n=x, y, z, w$). (b) shows the quasienergy spectra of the region $\varepsilon\in(0, \omega)$ in (a). The gray dashed line (red solid line) represents the amplitude of the time-periodic vector potential $A=0$ ($A=1.5$). (c) Band structure in the plane $(\varepsilon, k=k_{y}=k_{z}=k_{w})$ when the open boundary condition is along the $x$ direction. (d) Band structure in the quasienergy interval $\varepsilon\in(4.5, 6.5)$ in which there are gapless boundary states. In (c) and (d), the amplitude is $A=1.5$. Here, we choose $\omega=11$ and $m=-6$.}%
	\label{fig4}
\end{figure}

In this section, we apply a time-periodic vector potential ${\boldsymbol{A}}(\tau)$ to the 4D Dirac model $H(\textbf{k})$, then the time-dependent Hamiltonian $\mathcal{H}(\textbf{k}, \tau)$ is given by the following expression:
\begin{align}
\mathcal{H}(\textbf{k}, \tau)&=\sum_{j=1}^{5}h_{j}(\textbf{k}, \tau)\Gamma_{j},
\end{align}
with
\begin{align}
h_{1}(\textbf{k}, \tau)&=m+c\sum_{n=x, y, z, w}\cos[k_{n}+A\cos(\omega\tau)],\nonumber\\
h_{2}(\textbf{k}, \tau)&=\sin[k_{x}+A\cos(\omega\tau)],\nonumber\\
h_{3}(\textbf{k}, \tau)&=\sin[k_{y}+A\cos(\omega\tau)],\nonumber\\
h_{4}(\textbf{k}, \tau)&=\sin[k_{z}+A\cos(\omega\tau)],\nonumber\\
h_{5}(\textbf{k}, \tau)&=\sin[k_{w}+A\cos(\omega\tau)],
\end{align}
where $A$ is the amplitude of the time-periodic vector periodic. After the Fourier transformation, the block matrices in the Floquet Hamiltonian $\mathcal{H}_{F}$ are shown below:
\begin{align}
\mathcal{H}_{0}=&[\sin(k_{x})\Gamma_{2}+\sin(k_{y})\Gamma_{3}+\sin(k_{z})\Gamma_{4}\nonumber\\
&+\sin(k_{w})\Gamma_{5}]\mathcal{J}_{0}(A)+m(\textbf{k})\Gamma_{1},\\
\mathcal{H}_{-l}=&\sum_{j=1}^{5}h_{-l, j}\Gamma_{j}, \mathcal{H}_{+l}=\mathcal{H}_{-l}^{\dagger},
\end{align}
with
\begin{align}
h_{-l, 1}=&\frac{c}{2}\sum_{n=x,y,z,w}[e^{i k_{n}}+(-1)^{l}e^{-i k_{n}}]\mathcal{J}_{l}(A),\nonumber\\
h_{-l, 2}=&-\frac{i}{2}[e^{i k_{x}}-(-1)^{l}e^{-i k_{x}}]\mathcal{J}_{l}(A),\nonumber\\
h_{-l, 3}=&-\frac{i}{2}[e^{i k_{y}}-(-1)^{l}e^{-i k_{y}}]\mathcal{J}_{l}(A),\nonumber\\
h_{-l, 4}=&-\frac{i}{2}[e^{i k_{z}}-(-1)^{l}e^{-i k_{z}}]\mathcal{J}_{l}(A),\nonumber\\
h_{-l, 5}=&-\frac{i}{2}[e^{i k_{w}}-(-1)^{l}e^{-i k_{w}}]\mathcal{J}_{l}(A),
\end{align}
where $m(\textbf{k})=m+c[\cos(k_{x})+\cos(k_{y})+\cos(k_{z})+\cos(k_{w}) ]\mathcal{J}_{0}(A)$. $\mathcal{J}_{l}(A)$ is the $l$th Bessel function of the first kind.

\begin{figure}[t]
	\includegraphics[width=8.5cm]{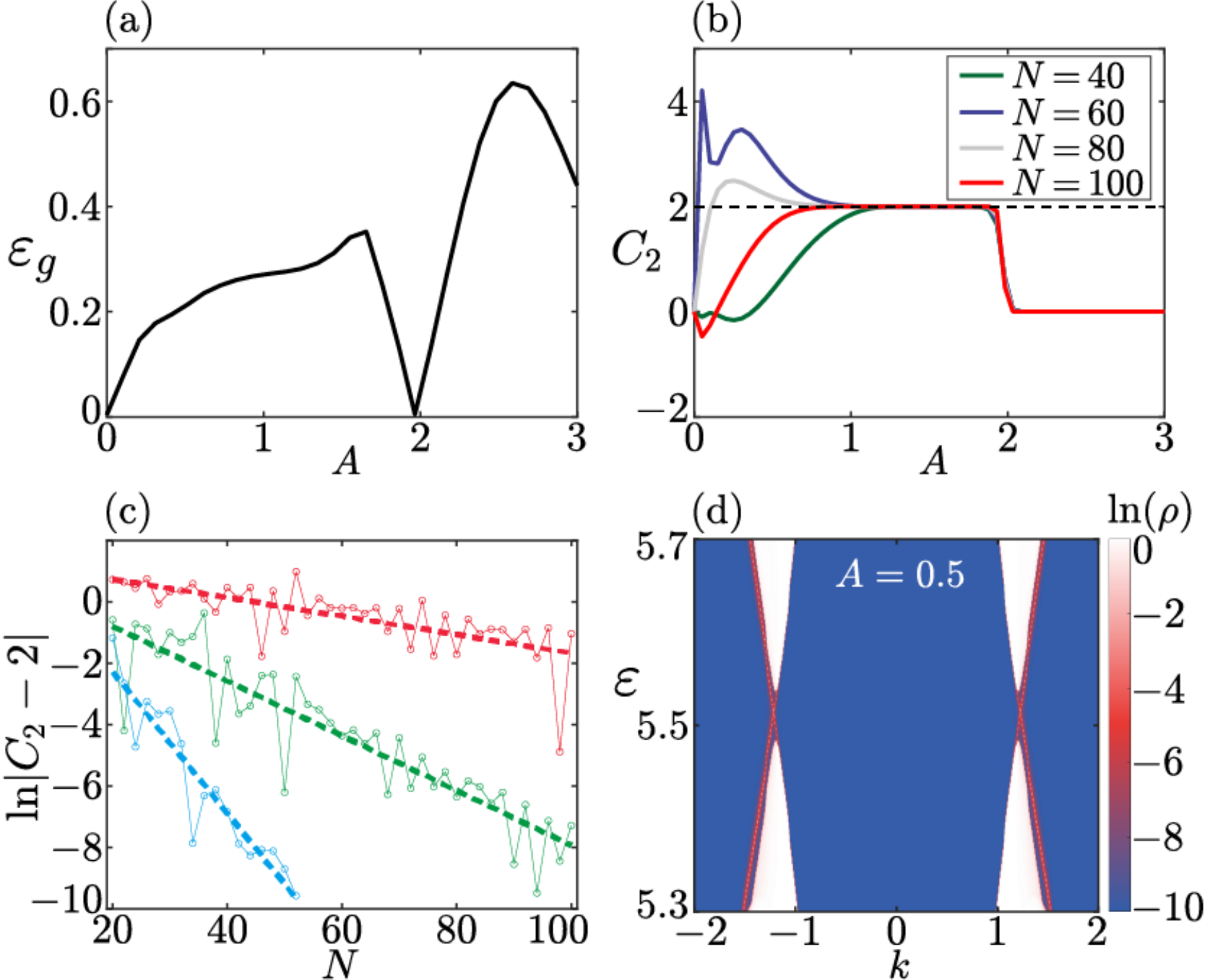} \caption{(a) The bulk gap $\varepsilon_{g}$ for the Floquet Hamiltonian $\mathcal{H}_{F}$ as a function of $A$. (b) The second Chern number $C_{2}$ as a function of $A$. The color of the solid line represents the number of $\textbf{k}$ points in the first Brillouin zone for the calculation of the second Chern number, which is $N^{4}$. (c) $\rm{ln}|C_{2}-2|$ as a function of $N$ for different amplitudes $A$. The red, green, and cyan dotted lines correspond to $A=0.5$, $A=1$, and $A=1.5$, respectively. The dashed lines of the linear distribution are obtained by fitting the dotted lines. (d) Band structure in the plane $(\varepsilon, k=k_{y}=k_{z}=k_{w})$ when the open boundary condition is along the $x$ direction. We choose the amplitude of the time-periodic vector potential $A=0.5$ in (d). Here, the parameters are $\omega=11$ and $m=-6$.}%
	\label{fig5}
\end{figure}

By solving the Floquet Hamiltonian $\mathcal{H}_{F}$, we present the bulk quasienergy spectra for $A=0$ and $A=1.5$ as shown in Figs.~\ref{fig4}(a) and \ref{fig4}(b), with the horizontal axis representing $k=k_{n}$ ($n=x, y, z, w$). The gray dashed line (red solid line) represents the amplitude of the time-periodic vector potential $A=0$ ($A=1.5$). When $A=0$, the Floquet Hamiltonian $\mathcal{H}_{F}$ contains only diagonal blocks, and the bands of different blocks overlap each other in the interval $\varepsilon\in\left ( 0, \omega \right ) $. After applying the time-periodic vector potential, the band gap in the interval $\varepsilon\in\left ( 0, \omega \right ) $ is opened as shown by the red solid line in Fig.~\ref{fig4}(b). In Figs.~\ref{fig4}(c) and \ref{fig4}(d) we show the band structure when the open boundary condition is along the $x$ direction. One can find that there are two gapless Dirac points in these gaps opened by the time-periodic vector potential, and they are distributed on the diagonal ($k=k_{y}=k_{z}=k_{w}$) in the first Brillouin zone of the quasi-3D system. We emphasize that these gapless boundary states appear only in the resonant quasienergy region $\varepsilon\in\left ( 0, \omega \right ) $ and are not present in the gap near $\varepsilon=0$.

In Fig.~\ref{fig5}(a), we show the bulk gap of the Floquet Hamiltonian $\mathcal{H}_{F}$ as a function of $A$. It is obvious that when $0<A<1.966$, the bulk gap is opened by the time-periodic vector potential, and there are gapless boundary states in this gap. In order to investigate the topological phase transition of the system, we show the evolution of the second Chern number $C_{2}$ with amplitude $A$ for different $N$ in Fig.~\ref{fig5}(b), where $N^{4}$ represents the number of $\textbf{k}$ points in the first Brillouin zone for calculating the second Chern number. In the interval $A\in(1,1.966)$, the second Chern number exhibits a nonzero quantized plateau $C_{2}=2$. When the bulk gap is closed at $A=1.966$, the second Chern number declines from a quantized value of two to zero. In addition, one can note that for weak amplitudes ($0<A<1$), the second Chern number exhibits nonzero fractional values. In Fig.~\ref{fig5}(c), we show the variation of the second Chern number $C_{2}$ with $N$. The value of the second Chern number oscillates toward $C_{2}=2$ as $N$ increases. We fit the evolution of $\rm{ln}|C_{2}-2|$ with $N$ as shown by the dashed line in Fig.~\ref{fig5}(c). It can be found that $\rm{ln}|C_{2}-2|$ decays linearly as $N$ increases, i.e., the value of the second Chern number approaches $C_{2}=2$ with an exponential trend. Figure \ref{fig5}(c) shows that the rate of $\rm{ln}|C_{2}-2|$ decay is related to the amplitude $A$. When the number of $\textbf{k}$ points $N$ in the discrete first Brillouin zone for calculation is sufficiently large, the second Chern number can approach the integer value $C_{2}=2$, and the number of $\textbf{k}$ points required for the calculation decreases as the amplitude $A$ is enhanced. In Fig.~\ref{fig5}(d), we show the boundary band structure when $A=0.5$, which implies that a weak vector potential can still induce the occurrence of topological boundary states.

\section{Conclusion}
\label{Conclusion}
In this paper, we present the scheme for inducing 4D FTI via two types of time-periodic driving, including time-periodic onsite potential ${\boldsymbol{V}}(\tau)$ and the time-periodic vector potential ${\boldsymbol{A}}(\tau)$. First, we introduce a time-periodic onsite potential into the 4D Dirac model. For the trivial static system, there is no boundary state in the bulk gap. When the frequency of the time-periodic onsite potential $\omega$ is less than the bandwidth of the static system $E_{W}$, the driven bands and the undriven bands overlap each other in the resonant quasienergy region $\varepsilon\in(0, \omega)$. We find that the time-periodic onsite potential can open the bulk gap in the resonance quasienergy region, leading to 3D gapless boundary states for the trivial static system. The system with 3D gapless boundary states is characterized by an emergent second Chern number $C_{2}$, referred to as a 4D FTI. By numerically calculating the evolution of the second Chern number with the amplitude of the time-periodic onsite potential $V$, it can be found that when the time-periodic onsite potential is introduced into the 4D system, the second Chern number exhibits a quantized plateau with $C_{2}=-3$. Moreover, when the amplitude $V$ increases to a critical value, the bulk gap closes and is accompanied by a phase transition of the system from a topological phase with $C_{2}=-3$ to another topological phase $C_{2}=3$. Additionally, we find that the topological phase of the 4D FTI can be modulated by tuning the frequency of the time-periodic onsite potential. The time-periodic onsite potential with different frequency can induce a topological phase transition from a 4D normal insulator to a 4D FTI with $C_{2}=-3$, $C_{2}=-1$, or $C_{2}=3$.

Second, we study the influence of the time-periodic vector potential ${\boldsymbol{A}}(\tau)$ in the 4D system. Similar to the time-periodic onsite potential, the time-periodic vector potential can also open the bulk gap in the resonant quasienergy region, and there are 3D gapless boundary states in this gap. By calculating the variation of the bulk gap and the second Chern number with the amplitude of the time-periodic vector potential $A$, we show that the vector potential can induce the emergence of a topologically nontrivial phase with $C_{2}=2$. Moreover, when the amplitude $A$ increases to a critical value, the bulk gap closes and is accompanied by a phase transition from the 4D FTI phase with $C_{2}=2$ to the trivial phase with $C_{2}=0$.

In this work, we confirm that the bulk-boundary correspondence for the 4D FTI is indeed captured by the second Chern number, and no boundary states are observed when the second Chern number is zero. The topologically protected Floquet boundary modes may appear in other 4D systems with a zero second Chern number~\cite{PhysRevX.3.031005}, which we will continue to investigate in future work.

Due to the limited dimensionality, 4D TIs are impossible to realize in real materials. However, in recent years, artificial systems have emerged as promising platforms for the experimental realization of 4D TIs. So far, 4D TIs have been realized in photonic crystals~\cite{PhysRevLett.108.133001, PhysRevLett.115.195303, PhysRevA.87.013814, PhysRevA.93.043827, 10.1038/nature25011, 10.1093/nsr/nwac289, 10.1515/nanoph-2022-0778}, acoustic lattices~\cite{PhysRevX.11.011016}, electric circuits~\cite{10.1038/s41467-020-15940-3, 10.1093/nsr/nwaa065, 10.1038/s41467-023-36359-6, 10.1038/s41467-023-36767-8}, and ultracold atom systems~\cite{PhysRevLett.115.195303}. Additionally, recent research indicates that Floquet systems can be realized in solid-state~\cite{doi:10.1126/science.1239834, 10.1038/s41567-019-0698-y, 10.1364/prj.404163}, photonic~\cite{10.1038/ncomms1872, 10.1038/nature12066, 10.1038/ncomms13756, 10.1038/ncomms13918, RevModPhys.91.015006, 10.1038/s41563-020-0641-8}, acoustic~\cite{10.1038/ncomms11744, 10.1038/ncomms13368}, electric circuits~\cite{dabiri2023electric}, and ultracold atom systems~\cite{10.1038/nature13915, 10.1038/s41567-019-0417-8, 10.1038/s41567-020-0949-y}. Therefore, we expect that the 4D FTI phase will soon be experimentally realized in artificial systems.

It is worth mentioning that in our other work~\cite{Liu2024Floquet4D}, we investigate the effects of high-frequency time-periodic driving in a 4D TI. In that work, the frequency of the time-periodic driving is greater than the bandwidth of the static system, and the driven bands do not overlap with the undriven bands. Therefore, we focus on the topological phase transition in the off-resonant quasienergy gap. It is found that the second Chern number of 4D TIs can be modulated by tuning the amplitude of time-periodic driving.

\section*{Acknowledgments}
B.Z. was supported by the NSFC (Grant No. 12074107), the program of outstanding young and middle-aged scientific and technological innovation team of colleges and universities in Hubei Province (Grant No. T2020001), and the innovation group project of the Natural Science Foundation of Hubei Province of China (Grant No. 2022CFA012). R.C. acknowledges the support of NSFC (under Grant No. 12304195) and the Chutian Scholars Program in Hubei Province. Z.-R.L. was supported by the Postdoctoral Fellowship Program of CPSF (under Grant No. GZC20230751) and the Postdoctoral Innovation Research Program in Hubei Province (under Grant No. 351342).

\bibliographystyle{apsrev4-1-etal-title_6authors}
%

\end{document}